\documentclass[aps,pra,twocolumn,superscriptaddress]{revtex4}

\usepackage[T1]{fontenc}
\usepackage[utf8]{inputenc}
\usepackage{graphicx}
\usepackage{dcolumn}
\usepackage{bm}
\usepackage{ulem}
\usepackage{xcolor}
\usepackage{lipsum}
\usepackage{hyperref}
\usepackage{amsmath}
\hypersetup{
    colorlinks=true,
    linkcolor=blue,
    filecolor=magenta,      
    urlcolor=black,
    citecolor=blue,
    pdftitle={Overleaf Example},
    pdfpagemode=FullScreen,
    }

\begin{document}

\preprint{APS/123-QED}

\title{Quantum-enhanced phase sensing via spectral noise reduction}

\author{Romain Dalidet$^1$, Anthony Martin$^1$, Audrey Dot$^2$, Inès Ghorbel$^2$, Loïc Morvan$^2$, Laurent Labonté}
\email{laurent.labonte@univ-cotedazur.fr}
\author{Sébastien Tanzilli}
\affiliation{Université Côte d’Azur, CNRS, Institut de physique de Nice (INPHYNI), France \\
Thales Research and Technology (TRT), Thales Group, Palaiseau, France}

\begin{abstract}
We report a direct demonstration of quantum-enhanced sensing in the Fourier domain by comparing single- and two-photon interference in a fiber-based interferometer under strictly identical noise conditions. The simultaneous acquisition of both signals provides a common-mode reference that enables a fair and unambiguous benchmark of quantum advantage. Spectral analysis of the interferometric outputs reveals that quantum correlations do not increase the amplitude of the modulation peak, but instead lower the associated noise floor, resulting in the expected 3~dB improvement in signal-to-noise ratio. This enhancement persists in the sub-shot-noise regime, where the classical signal becomes indistinguishable from the spectral background while the two-photon contribution remains resolvable. These observations establish quantum super-sensitivity in the Fourier-domain as an operational and broadly applicable resource for precision interferometric sensing.
 \end{abstract}

\maketitle

\section{Introduction}
Interferometric sensors, encompassing gravitational-wave detectors, distributed fiber systems and acoustic, optical probes, are usually operated and characterized in the frequency domain. Weak physical signals are identified as narrow spectral lines emerging from a continuous background in the power spectral density (PSD), rather than as absolute phase shifts in the time domain \cite{Fan2020,Zhu2023,Liao2025,Koyamada2009}. Their sensitivity is therefore determined by the spectral contrast, namely comparing the height of a signal peak to the noise floor in the Fourier space. In practice, detection thresholds and integration times are set by the measured noise spectra and by the resulting spectral signal-to-noise ratios \cite{He2021,Yang2021,Deepa2020}.

Spectral-domain analysis is widely used in both classical interferometry \cite{Fan2020,Zhu2023,Liao2025,Koyamada2009,He2021,Yang2021,Deepa2020} and quantum optics, for instance quantum state characterization \cite{Thiel2020}, spatial and spectral interference \cite{Jin2021,Shimizu2006}, acoustic-wave detection \cite{Kaiser2022}, coherence tomography \cite{Kolenderska2020}, as well as fractional Fourier-domain sensing \cite{Weimann2016,Niewelt2023,Hegde2025}. On the other hand, quantum-enhanced sensitivity is almost exclusively formulated in the time or phase domain, in terms of variances, Fisher information, and Cramér--Rao bounds \cite{Giovannetti2011,Pezze2018,Huang2024,Montenegro2025}. As a result, quantum advantage is usually benchmarked through phase estimation, while practical interferometric sensors are optimized and compared through their PSD. A key open question is whether quantum correlations induced by entanglement can provide a direct and operational advantage in the spectral domain by reducing the spectral noise floor against which signals are detected.

Here we demonstrate quantum enhancement directly obtained in the Fourier domain of an interferometric signal. By comparing single- and two-photon interference under strictly identical technical-noise conditions, we analyse the corresponding power spectral densities and show that quantum correlations do not modify the spectral amplitude of the signal peak, but reduce instead the noise background. This leads to an increased spectral signal-to-noise ratio and provides a direct, operational form of quantum-enhanced sensitivity in the frequency domain. Our results establish spectral noise reduction as a key resource for quantum-enhanced interferometric sensing, expressed with the same physical quantities than those used for characterizing and optimizing broadband, noise-limited sensors.

\section{Theoretical framework}

Since interferometric sensors are usually characterized in the frequency domain, we first formulate quantum-enhanced sensitivity in terms of the PSD associated with a Poissonian photon flux. For a photonic source showing a mean emission rate $\lambda$, detected in discrete time bins of duration $\Delta t = f_0^{-1}$, each bin contains a random number of counts $I_k$ with a mean value $\langle I_k\rangle=\lambda\Delta t$. The corresponding PSD, obtained from the discrete Fourier transform of the sequence $\{I_k\}$, is flat and reads
\begin{equation}
    S_{I_k}(f) = \frac{\lambda}{f_0^2},
\end{equation}
which defines the shot-noise-limited spectral background (see Appendix~\ref{Appendix A}). Expressed in units of $counts^2/\mathrm{Hz}$, this quantity represents the variance of photon-number fluctuations per unit bandwidth and thus sets the noise floor against which any spectral signal must be resolved.

\begin{figure}[!h]
    \centering
    \includegraphics[width=1\linewidth]{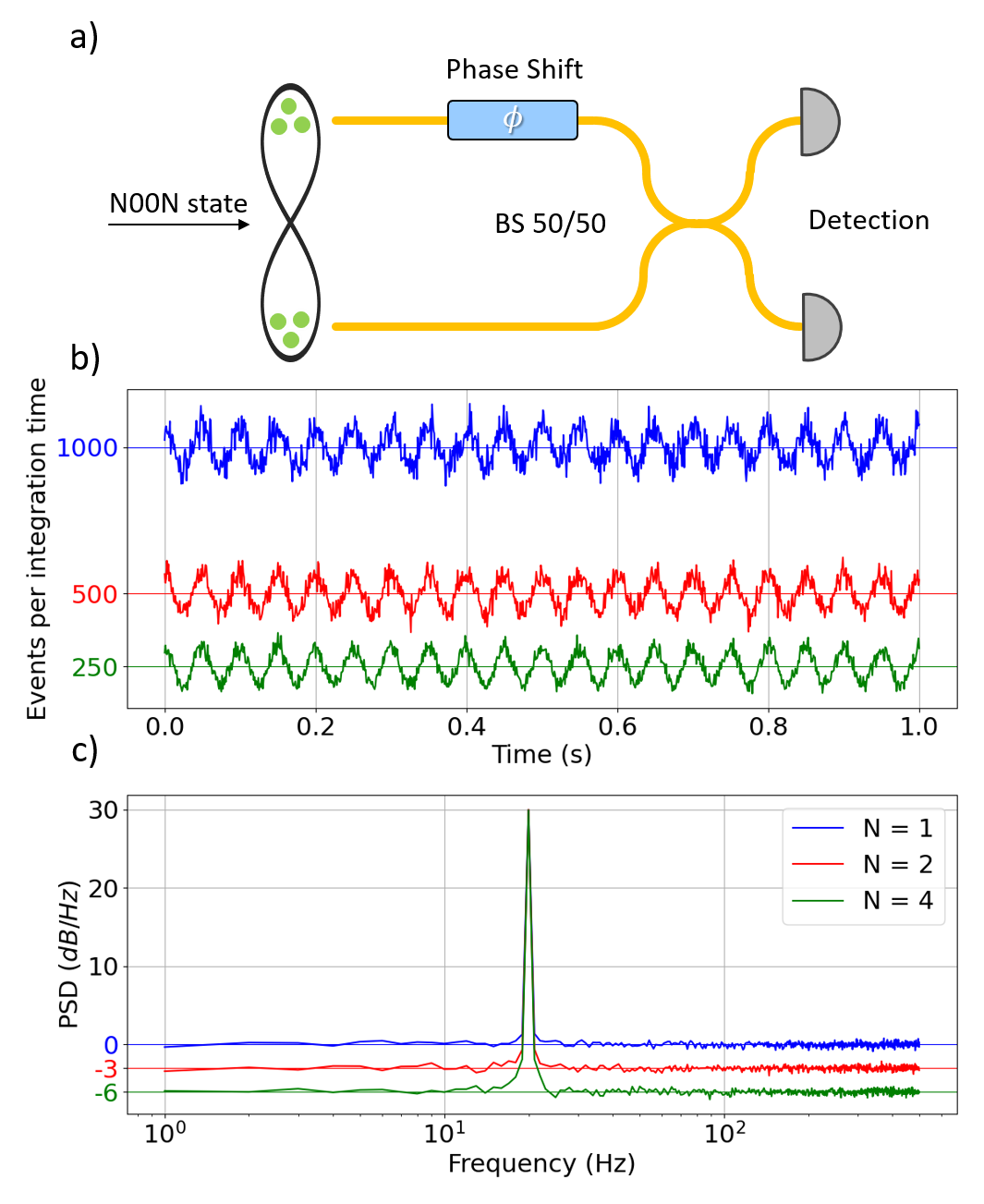}
\caption{(a) Output 50/50 beam splitter of the folded Franson interferometer where the two modes are recombined. The quantum probe, prepared in a N00N state, is coherently delocalized over the two arms, with the relative phase encoded in the upper arm. BS: beam splitter. 
(b) Numerical simulations of Eq.~(\ref{eq: probability N-coincidence}) for $N=1,2$, and $4$ photons (blue, red, and green curves, respectively), using $\lambda = 2\times10^{6}$ photons/s, $f_{0}=1$~kHz, $f_{m}=20$~Hz, and $A_{m}\simeq6.3\times10^{-2}$~rad. All interference fringes have the same amplitude, while the absolute noise associated with Poissonian statistics decreases with increasing $N$. (c) Corresponding power spectral densities computed from the temporal signals in (b). The spectral peak has the same amplitude (about $30$~dB/Hz) for all $N$, whereas the noise floor scales inversely with $N$.}

    \label{fig: theory}
    \end{figure}

We now consider an interferometric configuration in which the probe field is prepared in a correlated $N$-photon state and injected into a folded Franson interferometer \cite{Franson1989}. Fig. \ref{fig: theory}(a) depicts the output balanced beam splitter (BS) where the two modes are recombined. Here, the input state reads
\begin{equation}\label{eq: NOON state}
|\psi\rangle = \frac{|N,0\rangle_{ab} + e^{iN\phi}|0,N\rangle_{ab}}{\sqrt{2}},
\end{equation}
where $a$ and $b$ denote the two arms of the interferometer. Such path-entangled states constitute a canonical resource for quantum-enhanced interferometry \cite{Giovannetti2004,Dowling2008}. The $N$-fold phase factor reflects the collective response of the probe to the accumulated phase and is responsible for phase super-resolution. The probability to register an $N$-fold coincidence at the output reads
\begin{equation}\label{eq: probability N-coincidence}
    P_N(\phi) = \frac{\lambda}{2N}\!\left[1 \pm V_N \cos(N\phi)\right],
\end{equation}
where $V_N$ stands for the interference visibility. The prefactor $1/N$ ensures a fair comparison between probes of different photon number by fixing the total optical energy per measurement.

To connect with realistic sensing scenarios, we consider a small phase modulation of the form
\begin{equation}\label{eq: relative phase modulation}
\phi(t) = A_m \cos(2\pi f_m t) + \phi_0 ,
\end{equation}
where $A_m$ and $f_m$ are the modulation amplitude and frequency, respectively, such that $A_m \ll V_N$. $\phi_0$ is the operating point, set at mid-fringe, $\phi_0=\pi/2N$. In this linear regime, the detection probability can be expanded to first order as
\begin{equation}\label{eq: probability N-coincidence Taylor}
    P_N(t) \simeq \frac{\lambda}{2N} \pm \frac{\lambda}{2} V_N A_m \cos(2\pi f_m t).
\end{equation}
The corresponding PSD then takes the form
\begin{equation}\label{eq: PSD phase}
    S_N(f) \simeq \frac{1}{f_0^2}
    \left[
        \frac{\lambda}{2N}
        + \frac{(\lambda V_N A_m)^2}{16}\,\delta(f-f_m)
    \right].
\end{equation}
It consists of a flat noise floor, set by the quantum statistics of the detected probes, and a narrow spectral line at the modulation frequency. Importantly, while the height of the spectral peak is independent of $N$, the noise floor scales as $1/N$. Quantum correlations therefore do not amplify the signal in the Fourier space. Instead, they reduce the noise background on which it is detected.

Physically, the time-domain modulation $P_N(t)$ exhibits the same oscillation amplitude for all $N$, but the fluctuations associated with photon counting decrease as $\sqrt{N}$, as illustrated in Fig.~\ref{fig: theory}(b,c). For a visibility equal to 1, the signal-to-noise ratio of the spectral line is given by
\begin{equation}
    \mathrm{SNR}(N)=\frac{\lambda A_m^2}{8}\,N ,
\end{equation}
leading to the scaling $\mathrm{SNR}(N)/\mathrm{SNR}(1)=N$. The quantum advantage in the Fourier domain thus comes from a reduction of the spectral noise floor, not from an enhancement of the signal amplitude. This noise redistribution in the frequency space yields a direct operational meaning of quantum super-sensitivity: entangled probes reveal spectral features that remain hidden beneath the shot-noise background for classical light. Fig.~\ref{fig: theory} illustrates this mechanism for $N=1$ and for $N=2,4$ photon states, corresponding to classical and entangled probes, respectively.

\begin{figure*}[!t]
    \centering
    \includegraphics[width=1\linewidth]{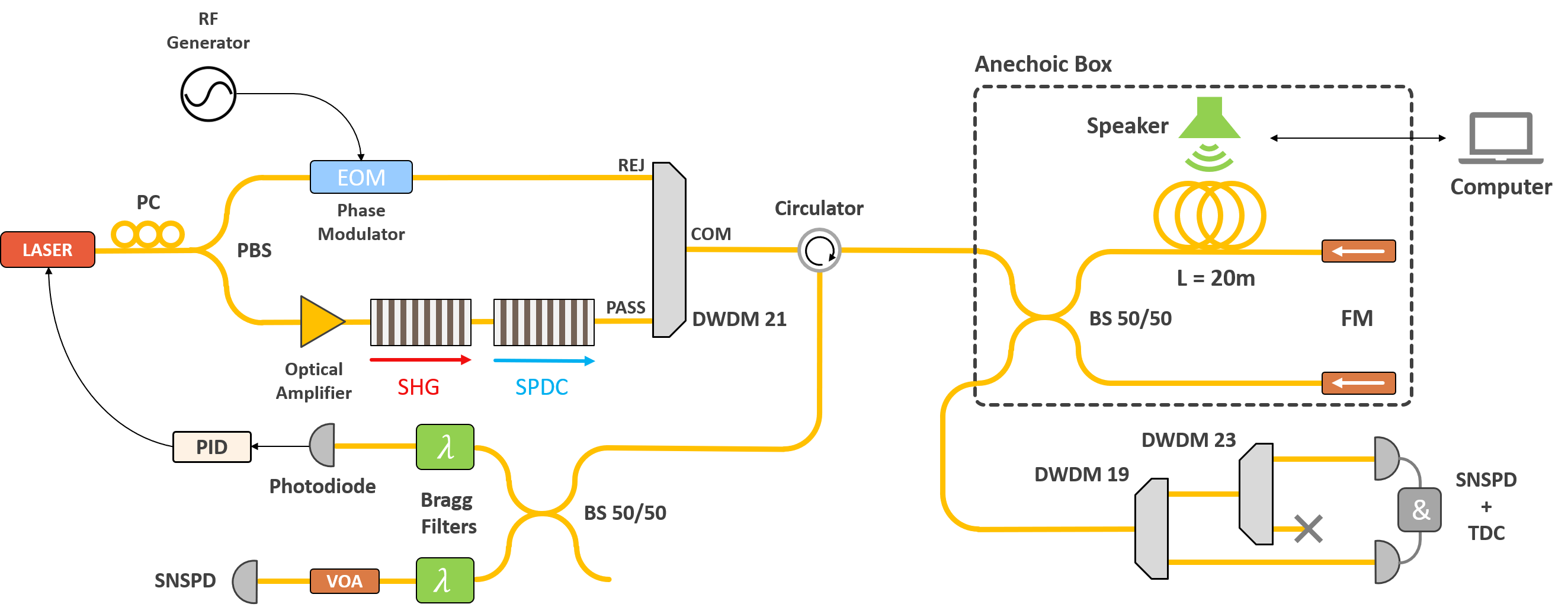}
    \caption{Experimental setup for demonstrating quantum-enhanced phase sensing in the frequency domain. A phase-modulated laser field and photon pairs are injected into a folded Franson interferometer incorporating a 20~m fiber-coil transducer. The interferometer outputs are analyzed using single-photon detection and two-photon coincidence measurements. EOM: electro-optic modulator; VOA: variable optical attenuator; BS: beam splitter; SNSPD: superconducting nanowire single-photon detector; TDC: time-to-digital converter; DWDM: wavelength demultiplexer with rejection (REJ), passband (PASS), and common (COM) ports.}

    \label{fig: experiment}
\end{figure*}

\section{Experimental implementation}

To test the above predictions in an operational setting, we directly compare single and two-photon interference within the same fiber-based setup. Both signals are recorded simultaneously, ensuring that they experience identical technical noise, including laser phase and intensity fluctuations as well as the intrinsic response of the interferometer. This simultaneous acquisition provides a rigorous and unbiased benchmark of quantum advantage in the Fourier domain, by isolating the effect of quantum correlations from all classical noise sources.

\textit{Apparatus.} The experimental setup is shown in Fig.~\ref{fig: experiment}. A continuous-wave telecom laser at $\lambda = 1560.61\,\mathrm{nm}$ (ITU channel~21) is split by a polarization controller and a polarizing beam splitter. One arm is sent through an electro-optic modulator (EOM), while the other is amplified and injected into two pigtailed type-0 lithium niobate waveguides, where second-harmonic generation followed by spontaneous parametric down-conversion produces time-correlated photon pairs. The generated pairs are combined with the pump using a dense wavelength-division multiplexer (DWDM, ITU channel~21) and, via an optical circulator, injected into a folded Franson interferometer in a Michelson configuration. The interferometer consists of a balanced beam splitter, a $20\,\mathrm{m}$ fiber coil acting as a phase-sensitive transducer and placed in front of a loudspeaker, and two Faraday mirrors providing intrinsic polarization compensation. Acoustic vibrations generated by the loudspeaker induce mechanical strain in the fiber coil, resulting in a modulation of the optical phase. The entire assembly is enclosed into a thermally stabilized anechoic chamber to suppress environmental perturbations.

The two output ports of the interferometer are analyzed in parallel. In one arm, cascaded DWDMs (ITU channels~19 and~23) deterministically separate the photons of each pair, reject the residual pump, and route them to superconducting nanowire single-photon detectors (SNSPDs) connected to a time-to-digital converter for coincidence measurements. In the other arm, a balanced beam splitter directs the light through $500\,\mathrm{MHz}$-bandwidth Bragg filters: one channel is detected with a classical photodiode, while the other passes through a variable optical attenuator before being detected by a SNSPD, enabling single-photon measurements under controlled flux conditions.

\textit{Measurement scheme.} The single- and two-photon interference signals are acquired simultaneously. We stress that they are both are stabilized at their respective mid-fringe operating points, where the phase-to-intensity response is maximal. Since both probes originate from the same laser, their working points cannot be independently adjusted by static phase shifts. This is achieved dynamically using the EOM. At the output of DWDM~21, the laser field in channel~21 is made of of a carrier frequency and two radio-frequency sidebands separated by the modulation frequency of the EOM ( $\sim10\,\mathrm{GHz}$). The adjacent channels contain the down-converted photon pairs. In the single-photon detection path, the upper Bragg filter selects the carrier, which is detected by the photodiode and used in a PID feedback loop to lock the interferometer at the mid-fringe point of the two-photon interference. The lower filter selects the $+1$ modulation sideband, whose relative phase can be finely tuned via the RF drive of the EOM. This independent control enables the single-photon interference to be simultaneously set at its optimal working point. It allows a truly synchronous and common-mode acquisition of classical and quantum interference in the Fourier domain, i.e., both signals are measured simultaneously under identical experimental conditions.

\section{Results}

The loudspeaker is driven at the reference frequency of $440\,\mathrm{Hz}$ (A4 musical tone), and the interferometric signals are sampled at $10\,\mathrm{kHz}$ over a $1\,\mathrm{s}$ acquisition window. By adjusting the optical amplification and the variable optical attenuator, the detected count rates are set to approximately $80\,\mathrm{kHz}$ for single-photon events and $40\,\mathrm{kHz}$ for two-photon coincidences. Independent calibrations yield interference visibilities of $V_1 = 99.3\%$ and $V_2 = 97.4\%$, limited respectively by detector dark counts ($\sim 100\,\mathrm{Hz}$) and by the finite coincidence-to-accidental ratio ($\sim 100$). For each acoustic power, the measurement is repeated 100 times in order to average down statistical fluctuations of the spectral noise floor.

\begin{figure}[h!]
    \centering
    \includegraphics[width=1\linewidth]{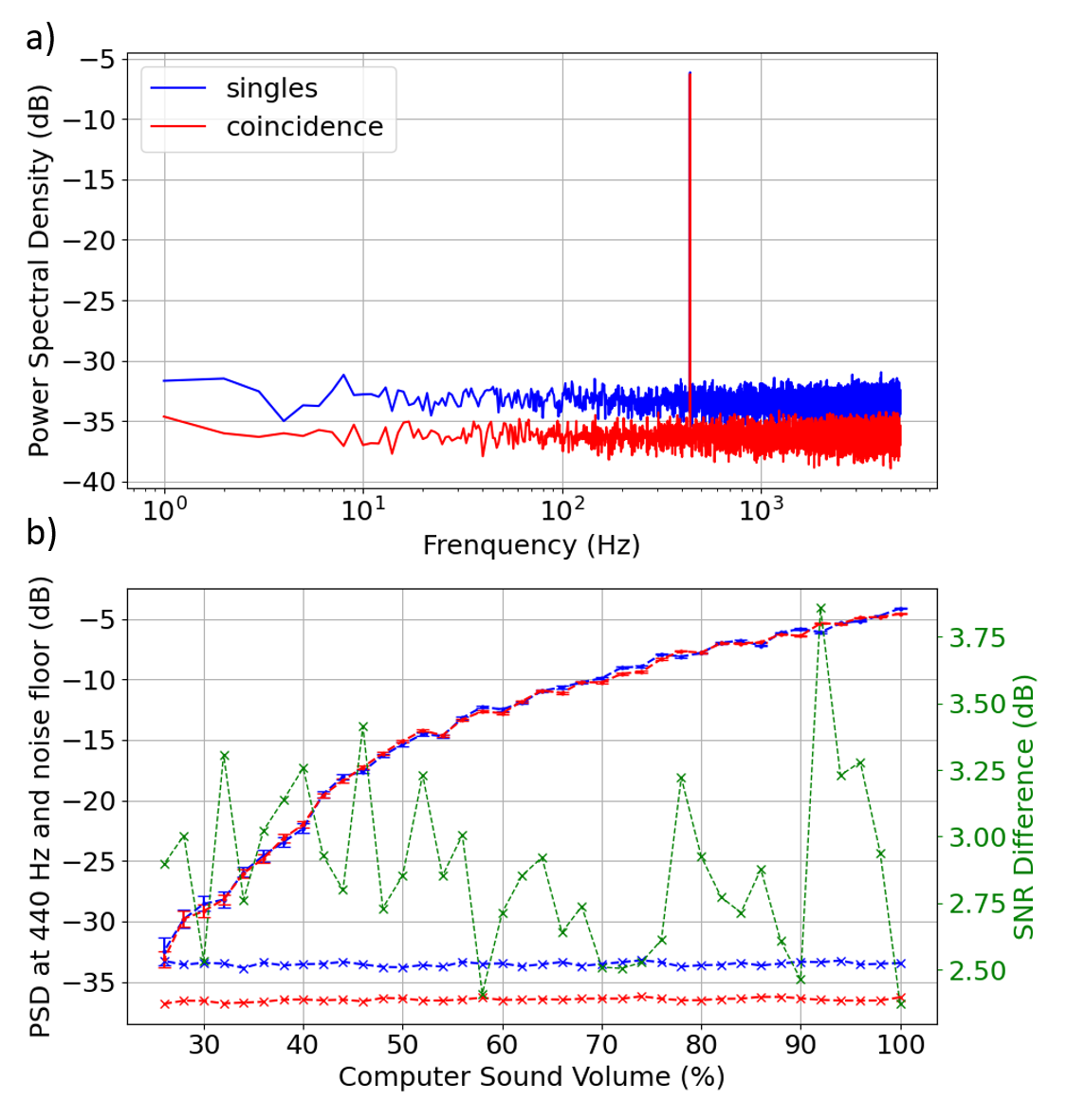}
    \caption{Power spectral densities (PSDs) of single- and two-photon interference at 440 Hz. (a) Example at 88\% sound volume showing identical spectral peaks but different noise floors. (b) Mean PSD values versus sound volume. Lower curves: noise floors separated by 3 dB. Upper curves: peak amplitudes at 440 Hz, identical for both probes. The error bars are calculated from the standard deviation of the 100 computed PSD per point. The green trace shows the measured SNR difference, consistent with the predicted 3 dB gain.}
    \label{fig: experimental results}
\end{figure}

A representative PSD recorded at $88\%$ of the maximum sound amplitude is shown in Fig.~\ref{fig: experimental results}(a), while Fig.~\ref{fig: experimental results}(b) summarizes the extracted spectral features as a function of the applied acoustic drive. The two lower traces correspond to the mean noise floors for single- and two-photon probes, which are separated by $3\,\mathrm{dB}$, in quantitative agreement with the expected reduction of the shot-noise level when the probe energy is distributed among correlated photon pairs. The two upper traces report the height of the spectral line at $440\,\mathrm{Hz}$, which is found to be identical in both cases. The green curve displays the measured difference in signal-to-noise ratio, clearly revealing the predicted $3\,\mathrm{dB}$ quantum enhancement in the Fourier domain.

\begin{figure}[h!]
    \centering
    \includegraphics[width=1\linewidth]{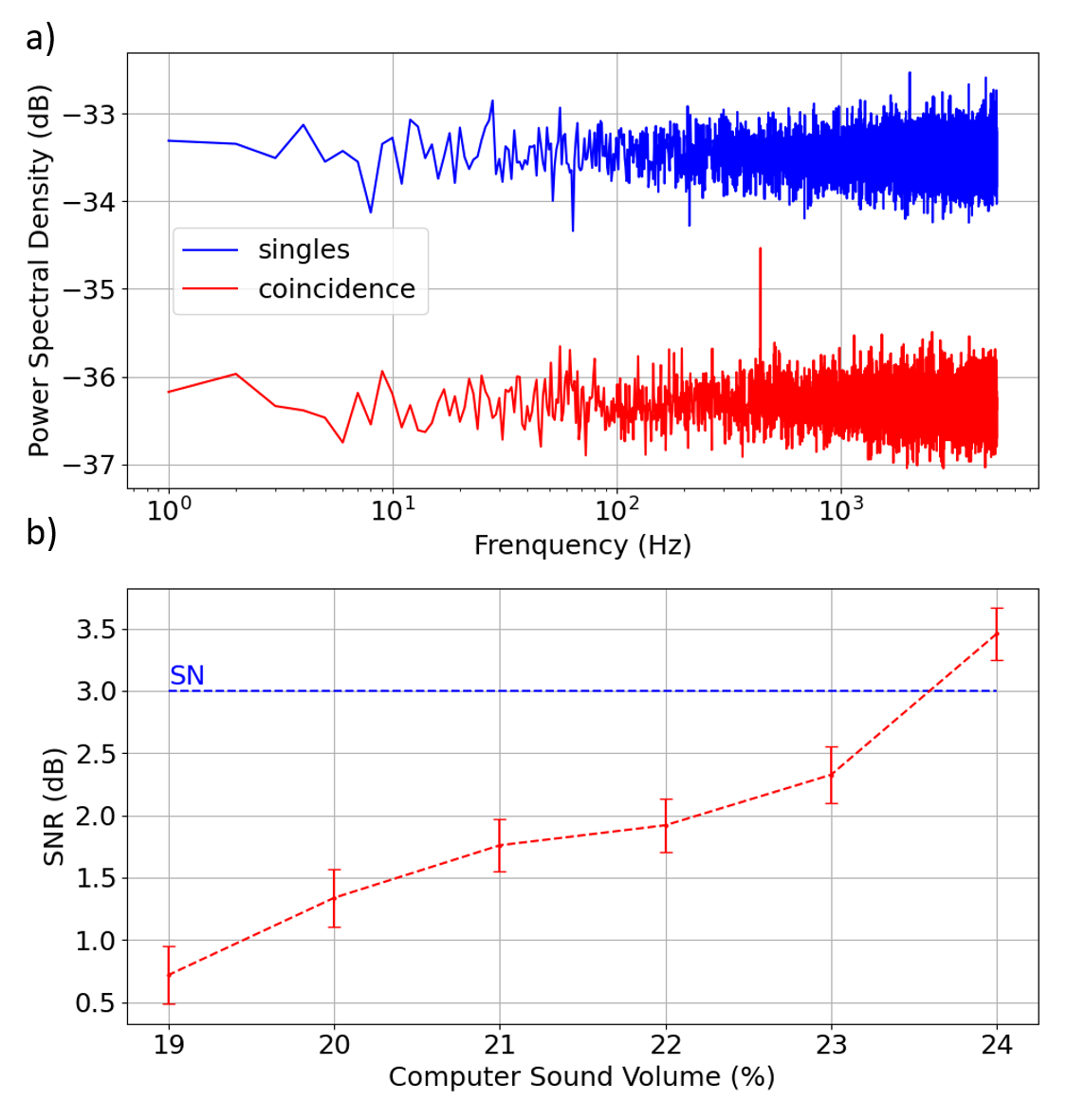}
    \caption{Two-photon performance in the low-signal regime. (a) PSD at 22\% sound volume, where the single-photon peak is lost in the noise. (b) Averaged results demonstrating sub-shot-noise sensitivity with two-photon probes, confirming quantum super-sensitivity in the Fourier domain. The errors bars are calculated from the standard deviation of the 1000 computed PSD per point. SN: shot-noise.}
    \label{fig: SNL measurement}
\end{figure}

As the acoustic amplitude is reduced, the single-photon spectral peak progressively merges with the shot-noise background and becomes undetectable below approximately 24\% of the maximum drive. To explore this low-signal regime, the acquisition is extended to 1000 realizations per point, further suppressing statistical fluctuations of the PSD. A typical spectrum at 22\% sound amplitude is shown in Fig.~\ref{fig: SNL measurement}(a), with the corresponding averaged results presented in Fig.~\ref{fig: SNL measurement}(b). While the classical peak is now indistinguishable from the noise floor, the two-photon contribution remains clearly resolved, demonstrating sub-shot-noise sensitivity. We emphasize that the shot-noise reference is defined by the number of detected photons and does not rely on any correction for global optical losses.

Beyond the specific demonstration with two-photon N00N states, our results highlight an alternative, conceptually different route to enhanced sensitivity, based on the reduction of the spectral noise floor rather than on an increase of the signal amplitude. In classical interferometry, improving the signal-to-noise ratio generally requires increasing the optical power, as dictated by shot-noise scaling, which is incompatible with photosensitive samples or with sensing scenarios operating under stringent photon-flux constraints \cite{Taylor2013,Wolfgramm2010,Crespi2012}. By contrast, the quantum enhancement demonstrated here arises from a reduction of the spectral noise floor at fixed probe power. This quantum spectral noise reduction therefore points to a direct advantage for low-light and non-invasive measurements, where a given spectral signal-to-noise ratio can be reached with fewer probe photons. An important open direction concerns the extension of this Fourier-domain approach to frequency regions dominated by colored technical noise, such as the ubiquitous $1/f$ noise limiting seismic, mechanical, and fiber-based sensors \cite{Abbott2016,Teixeira2014}. Understanding how quantum correlations modify not only the level but also the spectral structure of noise \cite{Collett1984,Clerk2010} could enable new strategies for separating signal and environmental fluctuations based on their distinct quantum statistical signatures.

\section{Conclusion}

We have demonstrated a quantum advantage that appears directly in the Fourier-domain description of an interferometric signal. By comparing single- and two-photon interference under identical technical-noise conditions, we showed that quantum correlations do not increase the amplitude of the spectral peak but instead reduce the underlying noise floor, resulting in a $3\,\mathrm{dB}$ improvement of the signal-to-noise ratio. This advantage persists even in a regime where the classical signal is buried in the spectral background while the quantum signal remains detectable, providing a direct experimental evidence of Fourier-domain quantum super-sensitivity.

Beyond validating the theoretical framework, these results establish spectral noise reduction as an operational quantum resource in the same representation used to characterize and optimize practical sensors. Since interferometric signals in fiber-based phase, acoustic, and distributed sensing systems are routinely processed and benchmarked through their PSD \cite{Teixeira2014,Gorshkov2022,Ghazali2024}, our work opens the route towards transferring quantum-enhanced sensitivity from laboratory demonstrations to realistic, broadband sensing platforms.

\section*{Data Availability}
Data are available from the authors on reasonable request.

\setcounter{section}{0}
\renewcommand{\thesection}{\Alph{section}}

\section{Derivation of the shot-noise limit via Fourier analysis}\label{Appendix A}

 Consider a basic photon detection setup with a source exhibiting Poissonian statistics (laser) with a mean emission and detection rate $\lambda$. For sake of simplicity, we assume an homogeneous statistics with $\lambda = C \in \mathbf{R}^+$ and an ideal, stationary detection process with instantaneous response. Thus, the measured signal is:

\begin{equation}\label{eq : signal poisson}
    I(t) = \sum_i \delta(t-t_i) \, ,
\end{equation}
where $t_i$ are the photon arrival times. By definition, using the Wiener–Khinchin theorem, the Power Spectral Density (PSD) of a continuous stochastic signal is given by :

\begin{equation}\label{eq : PSD autocor}
    S_I(f) = \hat{R}_I(f) \, ,
\end{equation}
where f is the frequency and $\hat{R}_I(f)$ is Fourier transform of the autocorrelation trace of the signal. As we are interested in the fluctuations of the signal, we calculate the autocorrelation trace of its zero-mean:
\begin{align}
    \tilde{I}(t) & = I(t) - \lambda \, ,\\
    \hat{R}_{\tilde{I}}(\tau)  & \triangleq \langle  \tilde{I}(t) \tilde{I}(t+\tau) \rangle = \lambda\delta(\tau) \, .
\end{align}
The Fourier transform of the above expression is straightforward and yields $S_{\tilde{I}}(f) = \lambda$. Consequently, photon counting from a Poissonian source results in a flat PSD equal to the average photon detection rate, defining the shot-noise limit. This result reflects the fundamental property of the Poisson process, namely, the absence of correlations in the photon arrival times $t_i$. \\
In a practical photon-counting experiment using single-photon detectors and a time-to-digital converter, the measurement process is discretized into time bins of width $\Delta t = f_0^{-1}$ where $f_0$ denotes the acquisition frequency. In this case, each $k$ time bin detects $I_k$ photons, defined by:
\begin{equation}\label{eq : Nb photon time bin}
    I_k = \int_{k\Delta t}^{(k+1)\Delta t}I(t)dt \, .
\end{equation}
From this sequence of time bins, we compute the discrete PSD (periodogram) via their discrete Fourier transform:
\begin{equation}\label{eq: periodogram}
    S_{I_k}(f) = \lim_{K\to\infty}\frac{1}{K} \left\langle \left| 
    \sum_{k=0}^{K-1} I_k e^{-2\pi i fk\Delta t} 
    \right|^2 \right\rangle \, .
\end{equation}
Since $I(t)$ is a Poissonian process, the photon counts in each time bin are statistically independent and follow a Poisson distribution with mean $\langle I_k\rangle=\lambda \Delta t$. Thus, the discrete PSD yields:
\begin{equation}
    S_{I_k}(f) = \frac{\lambda}{f_0^2} \, ,
\end{equation}
which correspond to the discrete shot-noise limit of the signal. The factor $f_0^2$ arises from the discrete nature of the measurement combined with the mean number of detected photon of each time bin. Recall that for a Poisson distribution, the mean equals the variance; thus, the units of the PSD, expressed as $counts^2/Hz$, correspond to the variance of the binned photon counts per unit bandwidth, $\sigma^2/Hz$.

\end{document}